\documentclass[journal]{IEEEtran}
\usepackage{graphicx}
\usepackage{amsfonts}
\usepackage{amsmath}
\usepackage{boldline}
\usepackage{amsthm}
\usepackage{epstopdf}
\usepackage{amssymb}
\usepackage{tabu}
\usepackage{multirow}
\usepackage{multicol}
\usepackage{booktabs}
\usepackage{setspace}
\usepackage{algpseudocode}
\usepackage{algcompatible}
\usepackage{mathrsfs,dsfont}
\usepackage{color}
\usepackage{float}
\usepackage{enumitem}
\usepackage{epsfig,cite}
\usepackage{url}
\usepackage{cases}
\usepackage{mathtools}
\usepackage{stackengine}
\usepackage[caption = false]{subfig}
\usepackage[linesnumbered,lined,boxed,commentsnumbered,ruled,longend]{algorithm2e}
\setlist[itemize]{leftmargin=*}
\newcommand\xrowht[2][0]{\addstackgap[.5\dimexpr#2\relax]{\vphantom{#1}}}
\newcommand{\tabincell}[2]{\begin{tabular}{@{}#1@{}}#2\end{tabular}}

\DeclarePairedDelimiter\norm{\lVert}{\rVert}%


\expandafter\def\expandafter\normalsize\expandafter{%
 	\normalsize
 	\setlength\abovedisplayskip{0.5pt}
 	\setlength\belowdisplayskip{0.5pt}
 	\setlength\abovedisplayshortskip{0.5pt}
 	\setlength\belowdisplayshortskip{0.5pt}
}

\usepackage{siunitx} 
\usepackage[american]{circuitikz} 
\usepackage{adjustbox}
\usepackage{mathtools}
\usepackage{tikz}
\usetikzlibrary{shapes,arrows,positioning,calc}

\usetikzlibrary{arrows}
\usepackage{verbatim}
\usepackage{tikz}
\usetikzlibrary{arrows}
\usepackage{verbatim}
\usetikzlibrary{shapes, arrows,positioning}
\usetikzlibrary{arrows.meta,positioning,fit,calc}
\usepackage{tikz}
\usetikzlibrary{shapes,arrows.meta,decorations}
\pgfdeclaredecoration{strange pin}{initial}
{
\state{initial}[width=0pt,next state=final] {
    \pgfpathlineto{\pgfpointadd{\pgfpointdecoratedpathlast}{\pgfpoint{0pt}{-2pt}}}
  }
  \state{final}
  {
    \pgfpathlineto{\pgfpointadd{\pgfpointdecoratedpathlast}{\pgfpoint{0pt}{-12pt}}}
  }
}
\tikzset{block/.style={draw, fill=blue!15, rectangle, 
    minimum height=2em, minimum width=4em, font = \LARGE},
    bl/.style={draw, fill=blue!15, rectangle, rotate=90,
    minimum height=2em, minimum width=4em, font = \LARGE}
sum/.style={draw, fill=blue!15, circle, node distance=1cm},
input/.style={coordinate},
output/.style={coordinate},
dot/.style = {circle,fill, inner sep=0.01pt, node contents={}},
       alr/.style = {Stealth-Stealth},
       arr/.style = {-Stealth},
HVgate/.style = {signal, draw, fill=blue!15, signal to=east,
                     font=\large\linespread{0.2}\selectfont, align=left},
custom pin/.style={pin edge={solid,thick,black,decorate,decoration={strange pin}}}}

\begin{document}
\vspace{-2.5cm}
\title{Measurement-Based Parameter Identification of DC-DC Converters with Adaptive Approximate Bayesian Computation}
\vspace{-2.5cm}


\normalsize{
\author{Seyyed Rashid Khazeiynasab and Issa Batarseh \\
Department of Electrical and Computer Engineering\\
University of Central Florida\\
Orlando, FL 32816 USA\\
Emails: rashid@knights.ucf.edu}
}

\maketitle

\begin{abstract}
The recent advances in power plants and energy resources have extended the applications of DC-DC converters in the power systems (especially in the context of DC micro-grids). Parameter identification can extract the parameters of the converters and generate accurate discrete simulation models. 
In this paper, we propose a measurement-based converter parameter calibration method by an adaptive Approximate Bayesian Computation with sequential Monte Carlo sampler (ABC SMC), which estimates the parameters
related to passive and parasitic components. At first, we propose to find suitable prior distribution for the parameter which we do not know the prior information about them. With having prior distributions, we can use the ABC-SMC to find the exact values of the parameters of the converter. We chose the distance function carefully and based on the simulations we assigned the best method for the threshold sequencing. For improving the computationally of the algorithm, we propose an adaptive weight that helps the algorithm to find the optimal values with fewer simulations. The effectiveness of the proposed method is validated for a DC-DC buck converter. The results show that the proposed approach can accurately and efficiently estimate the posterior distributions of the buck parameters. The proposed algorithm can be applied to other parameter identifications and optimization applications such as rectifiers, filters, etc.

\end{abstract}

\vspace{-0.5cm}
\section{Introduction}

Switch-mode power converters (SMPC) are broadly used in different power electronics applications, including motor
drives, computers, portable electronics, domestic appliances, or in power conversion systems
for renewable generation, among others \cite{ batarseh2018power}. Monitoring the conditions of the SMPCs and analyzing their outputs in the system plays an important role in the operation and reliability of power system. Estimating the parameters of SMPC can improve the mathematical models of the converters which are based on the linear analysis \cite{hayerikhiyavi2021gyrator}. Also for designing to design a good controller, the exact model of the SMPC is needed, which  relies on the exact parameters. On the other hand, the parameters of the SMPCs change with age, manufacturing tolerance, parasitic elements, and load changes. Consequently, these uncertainties must be
considered during the modeling stage of the power converter. For example, it has
been reported in \cite{ yang2010condition} that capacitors cause 30\% of the 
failures in converter circuits.  Failure of the converter or other power electronics component can even cause blackouts in the power system \cite{rcas, hayerikhiyavi2021comprehensive}. 
Then, the conditions of the SMPCs should be monitored, and converters' parameters for having an exact model should be estimated \cite{abu2000generalized}. 

In general, there are two categories of the system identification
technique; online and offline system identification
\cite{khazeiynasabpmu}. Since the model parameters depend on the operating conditions, the offline methods can not estimate the exact values of the parameters. In the online methods, real-time data or simulation-based data are obtained and used to identify the parameters of the system \cite{isgt}. 

The white-box based method in \cite{balakrishnan2018dc} has good accuracy, but its computational time is high, and for a complex system, its implementation is a big problem. The proposed methods in \cite{buiatti2007unified, rojas2020nonlinear} use a polynomial interpolation method
with the least-squares (LS) algorithm  to estimate the
parameters of the converter, but these methods can not find the global optimal, and under  different load changes,  the estimated parameters may be  different from the true parameters. The subspace-based method proposed in \cite{algreer2012microprocessor} has good accuracy, but its final solution needs heavy difficulties to implement.  In \cite{alonge2008identification, valdivia2009simple} the parameters of the DC-DC converter are estimated based on the (LS) technique, and finally, a non-linear black-box model of the converter proposed. But, in these methods, the physical parameters do not have meaning, and based on these types of models, we can not analyze the model correctly. The measurement-based approaches which are based on acquiring the instantaneous values of the input and output at the terminals of the power converters can be applied for parameter estimation \cite{rojas2020nonlinear}. These approaches have an advantage being compatible with
non-invasive online monitoring of the input/output signals. Then,
using these types of methods do not to conflict with the operation of the converters.

To overcome the aforementioned drawbacks, in this paper, we propose a measurement-based adaptive ABC-SMC method for converter parameter calibration and improve the efficiency of ABC-SMC by modifying the weight and assign an adaptive weight probability for the particles at each iteration. The contributions are summarized as follows. 
\begin{enumerate} 
\item We introduce a method to find the prior distribution of parameters that have unknown initial values, e.g., the impedance of the DC power supply, we introduce a method to find the prior distribution of the parameters without known the initial value. 
 \item We perform DC-DC converter parameter calibration by adaptive ABC SMC, which estimates the posterior distributions of the parameters by a simulation-based procedure. 
\item We improve the computational efficiency of ABC-SMC based parameter calibration by developing adaptive weights for particles at each iteration. This weighting scheme, helps the algorithm to avoid getting stuck in the local optimal. Also the algorithm needs less number of simulation to find the posterior distributions of the parameters.   
\end{enumerate}


The remainder of this paper is organized as follows. In Section \ref{sec:model}, we introduce the model under study.  Section \ref{sen} describes the sensitivity-based approach for identifying the most identifiable parameters. In Section \ref{ABC-general}, an overview is provided for ABC-SMC for parameter estimation. Section \ref{sec:proposed_abc} proposes an adaptive ABC-SMC approach. 
Section \ref{result} presents case studies to validate the effectiveness of the proposed method. 
Finally, conclusions are drawn in Section \ref{conclusion}. 

\section{DC-DC Buck Converter Model}\label{sec:model}
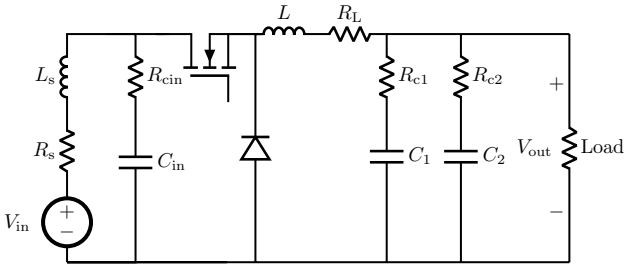
\begin{figure}[htb]
\centering
\begin{adjustbox}{max width=0.46\textwidth}
\centering
\begin{circuitikz} [line width=1pt][scale=1.1][/tikz/circuitikz/bipoles/length=0.9cm]
\ctikzset{resistors/scale=0.6, capacitors/scale=0.7, inductors/scale= 0.6,diodes/scale = 0.7, transistors/scale=1.2 }
\ctikzset{resistors/thickness=1.3, capacitors/thickness=1.3, inductors/thickness=1.3,diodes/thickness=1.3, transistors/thickness=1.3 }
\draw (0,0)
to[V,v=$V_\mathrm{in}$, invert] (0,1.4)
to[R,l=$R_\mathrm{s}$] (0,2.5)
to[L,l=$L_\mathrm{s}$] (0,4)
to[] (1.2,4)
to[R, l = $R_\mathrm{cin}$] (1.2,2.5)
to[C, l = $C_\mathrm{in}$] (1.2,1)
to[] (1.2,0)
to[] (0,0)
to[] (2.8,0)
(0.7,4) to [] (1.8,4)
(1.7,4) to[Tnigfete,n=mosfet, invert] (3.3,4)
to [diode, invert] (3.3, 0)
(3.3, 4) to [L, l =$L$] (4.3, 4)
 to [R, l =$R_\mathrm{L}$] (5.6,4 )
to[R, l = $R_\mathrm{c1}$] (5.6,2.5)
to[C, l = $C_\mathrm{1}$] (5.6,1.2)
(5.6,4 ) to [] (8.8, 4)
(6.9, 4) to[R, l = $R_\mathrm{c2}$] (6.9,2.5)
to[C, l = $C_\mathrm{2}$] (6.9,1.2)
(6.9,1.2) to [] (6.9,0)
(5.6,1.2) to [] (5.6, 0)
(8.8, 0) to [] (0,0)
(8.8, 4) to [ R,  l= $\mathrm{Load}$,  v=$V_\mathrm{out}$ ] (8.8, 0)
;
\end{circuitikz}
\end{adjustbox}
 \captionsetup{justification=raggedright,singlelinecheck=false}
 \caption{Circuit representation of the DC-DC buck converter.}\label{fig:buck}
\end{figure}

A buck converter is a form of DC to DC converter that can take input directly from a DC source, such as a battery.
Since the frequency of switching is high, then the parasitic elements of the converter component should be considered. We consider $R_\mathrm{M}$ as the the parasitic resistance for the MOSFET, $R_\mathrm{L}$ for inductor, $R_\mathrm{c}$ for the capacitor. We also model the input capacitor as a $C_\mathrm{in}$ series with a parasitic resistance as $R_\mathrm{cin}$ \cite{riba2018parameter}. The parasite element of the voltage source is also very important, we consider inductor as $L_\mathrm{s}$ series with resistance as $R_\mathrm{s}$.
Fig. \ref{fig:buck} shows the circuit of a DC-DC buck converter. 

As shown in Fig. \ref{fig:buck} the buck Converter circuit consists of the switching transistor, together with the flywheel circuit ($D$, $L$, and $C$). While the transistor is on, the current is flowing through the load via the inductor $L$. An inductor operates by opposing changes in current flow while also storing energy.  When the switching transistor is switched on, it is supplying the load with the current. The magnetic field around $L$ becomes released once the transistor switches off, releasing the energy that was stored in it. As a result, the voltage across the inductor is now in reverse polarity to that across $L$ during the 'on' period \cite{batarseh1993generalized}. 
The ration between the output voltage and input voltage of a buck converter can be written as follow:
\begin{align}
    \frac{V_\mathrm{out}}{V_\mathrm{in}} = D, \notag 
\end{align}
where $D$ is the duty cycle which is the ratio of the time which the switch is on to the whole time. 

\section{Identifying Critical Parameters} \label{sen}
\vspace{0.2cm}
After a model deficiency 
has been revealed, the next step is to identify the problematic parameters. A converter model with its control can have many parameters. Calibrating all parameters could be computationally challenging and also not every parameter is identifiable. Trajectory sensitivity has been used to identify the most critical parameters \cite{rpower}. 
Based on trajectory sensitivities, we can have insights about how the changes in parameters influence the system response. If a parameter exerts a large influence on the response, the corresponding sensitivity will be large. 

Specifically, the sensitivity of the outputs  $\boldsymbol{V}_\mathrm{out}$,  $\boldsymbol{I}_\mathrm{out}$ with regard to parameter $\alpha_i$ can be calculated as: 
\begin{align}
&S(\alpha_i) = \notag \\ 
&\sum_{k=1}^{K}{\frac{|V_{\mathrm{out}, k}(\alpha_i^+)\!-\!V_{\mathrm{out}, k}(\alpha_i^-)| +| I_{\mathrm{out}, k}(\alpha_i^+)\!-\!I_{\mathrm{out}, k}(\alpha_i^-)|}{K(\alpha_i^+\!-\!\alpha_i^-)/\alpha_i}}, \notag
\end{align}
where  $K$ is the number of time steps, $\alpha_i^+=\alpha_i + \Delta \alpha_i$ and $\alpha_i^-=\alpha_i-\Delta \alpha_i$, and $\Delta \alpha_i$ is a small perturbation of $\alpha_i$. 
After the sensitivity analysis, the parameters selected to be estimated are those with a large sensitivity \cite{khazeiynasabpmu, rpower}.

\section{DC-DC Converters Parameter Calibration by Adaptive ABC SMC} \label{ABC-general}
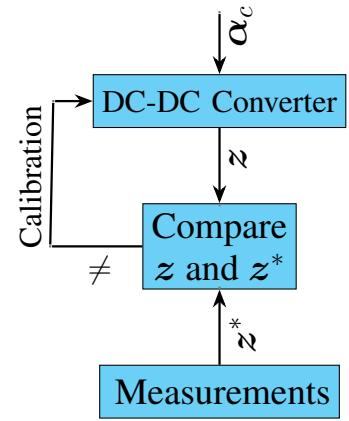
\begin{figure}[!htb]
\centering
\begin{adjustbox}{max width=0.3\textwidth}
\captionsetup{justification=raggedright,singlelinecheck=false}
\begin{tikzpicture}[
    font=\sf \scriptsize,
    >=LaTeX,
    cell/.style={rectangle, rounded corners=5mm, fill=green!15, draw,very thick,},
    operator/.style={circle,draw,inner sep=-0.5pt,minimum height =0.5cm, fill=red!10, font = \large}, 
    block/.style={draw, fill=cyan!50, rectangle, 
    minimum height=2em, minimum width=4em, font = \Large},
    blockv/.style={draw, fill= cyan!50, rectangle, font = \large,
    minimum height=2em, minimum width=4em},
    blockvv/.style={draw, fill= green!10, rectangle,
    minimum height=2em, minimum width=3em},
    function/.style={ellipse, draw,inner sep=1pt},
    ct/.style={circle,draw,line width = .75pt,minimum width=1cm,inner sep=1pt,},
    gt/.style={rectangle,draw,minimum width=4mm,minimum height=3mm,inner sep=1pt},
    dot/.style = {circle,fill, inner sep=0.001mm, fill=black!50, node contents={}},
    dotn/.style = {circle,fill, inner sep=0.0001mm, fill=white!100, node contents={}},
    dots/.style = {circle,fill, inner sep=0.1mm, fill=black!200, node contents={}},
    mylabel/.style={font=\scriptsize\sffamily},
    alr/.style = {-Stealth, dotted},
    dotss/.style = {dotted},
       arr/.style = {-Stealth, font = \large},
    ]
\node[block, name = dc, font=\large] {DC-DC Converter};
\node[block, below = 1cm of dc ] (sim) { \shortstack{Compare \\$\boldsymbol{z}$ and $\boldsymbol{z}^*$} };
\node[block, below = 1cm of sim ] (meas) {\,Measurements};
\node (dot1) [dot,above= 0.8cm of dc]{};
\node (dot2) [dot,left = 0.5cm of dc]{};
\node (dot3) [dot,below = 1cm of dc]{};
\node (dot3) [dot,above = 1cm of meas]{};
\node (dot5) [dot,left = 1.2cm of sim]{};
\draw[arr, line width=0.30mm] (dot1) -- node[pos = 0.205, right,  font = \Large]{\rotatebox{90}{$\boldsymbol{\alpha}_{c}$}}(dc);  
\draw[arr, line width=0.30mm] (dc) -- node[pos = 0.405, right,  font = \Large]{\rotatebox{90}{$\boldsymbol{z}$}}(sim);  
\draw[arr, line width=0.30mm] (meas) -- node[pos = 0.405, right,  font = \Large]{\rotatebox{90}{$\boldsymbol{z}^*$}}(sim);  
\draw[ line width=0.30mm] (dot5) -- node[pos = 0.48, left, font= \large]{ \rotatebox{90}{ Calibration }}(dot2);
\draw[arr, line width=0.30mm] (dot2) --node{}(dc);
\draw[ line width=0.30mm] (dot5) --node[pos = 0.53, below, font= \large]{$\neq$}(sim);
\end{tikzpicture}
\end{adjustbox}
 \captionsetup{justification=raggedright,singlelinecheck=false}
 \caption{The framework for DC-DC converter parameter estimation.}\label{fig:framework}
\end{figure}

Mathematical models have become powerful tools for model analysis. However, as the models become more complex, the computational challenges of parameter inference and model validation are increasingly vast. 
Let $\boldsymbol{z}^* = [\boldsymbol{V}_\mathrm{meas}^\top \;  \boldsymbol{I}_\mathrm{meas}^\top]^\top$ be the measurements with the actual value of the parameters, and $\boldsymbol{z} = [\boldsymbol{V}_\mathrm{out}^\top \;  \boldsymbol{I}_\mathrm{out}^\top]^\top$ be the outputs of model and $\boldsymbol{\alpha}_\mathrm{c}$ is the parameter vector which we want to estimate.  Assuming the prior distribution for $\boldsymbol{\alpha}_\mathrm{c}$ as $\pi(\boldsymbol{\alpha}_\mathrm{c})$. Fig. \ref{fig:framework} shows the framework for DC-DC converter parameter estimation.

ABC-based methods use systematic comparisons between real and simulated data in order to 
obtain a good approximation to the true (but unobtainable) posterior distribution
\begin{align}
    p(\boldsymbol{\alpha}_\mathrm{c} \vert \boldsymbol{z}^*) = \frac{ l(\boldsymbol{z}^* \vert \boldsymbol{\alpha}_\mathrm{c}) \pi(\boldsymbol{\alpha}_\mathrm{c})}{\int l(\boldsymbol{z}^* \vert \boldsymbol{\alpha}_\mathrm{c}) \pi(\boldsymbol{\alpha}_\mathrm{c})d\boldsymbol{\alpha}_\mathrm{c} }, 
\end{align}
where the denominator is referred to as the Bayesian Evidence; and the integral runs over all possible parameter values. $p(\boldsymbol{\alpha}_\mathrm{c}\vert \boldsymbol{z}^*)$ is posterior distribution, and $l(\boldsymbol{z}^* \vert \boldsymbol{\alpha}_\mathrm{c})$ is the likelihood of $\boldsymbol{\alpha}_\mathrm{c}$ given data $\boldsymbol{z}^*$. Instead of evaluating the likelihood, ABC-based approaches use systematic comparisons between real and simulated data.  
ABC  samples the prior and compares the simulated data $\boldsymbol{z}$ with the real data $\boldsymbol{z}^*$ and accepts only the simulations for which the distance between $\boldsymbol{z}^*$ and $\boldsymbol{z}$, $\rho(\boldsymbol{z},\boldsymbol{z}^*)$, is less than a predefined tolerance $\epsilon$. Therefore, the ABC algorithm provides
 the approximate posterior as \cite{silk2012optimizing}:
\begin{align}
    & p_{\epsilon}(\boldsymbol{\alpha}_\mathrm{c}\vert \boldsymbol{z}^*) 
     \propto  \int l(\boldsymbol{z} \vert \boldsymbol{\alpha}_\mathrm{c})  \mathds{1}  \big(\rho(\boldsymbol{z}, \boldsymbol{z}^*)\leq \epsilon \big) \pi(\boldsymbol{\alpha}_\mathrm{c}) d\boldsymbol{z},
\end{align}
where $p_{\epsilon}(\boldsymbol{\alpha}_\mathrm{c}\vert \boldsymbol{z}^*)$ is an approximation of the posterior $p(\boldsymbol{\alpha}_\mathrm{c}\vert \boldsymbol{z}^*)$ and $\mathds{1}(x)$ is equal to one when condition $x$ is true. \\
If $\epsilon$ is sufficiently small, the distribution $p_{\epsilon}(\boldsymbol{\alpha}_\mathrm{c}\vert \boldsymbol{z}^*)$ will be a good approximation of the posterior distribution. 
Recently, algorithms using Sequential
Monte Carlo (SMC) with particle filtering have gained growing attention \cite{beaumont2009adaptive, smart}.
ABC-SMC samples from a sequence of distributions that increasingly resemble the target posterior. 
They are constructed by estimating the intermediate distributions $p_{\epsilon_t}(\boldsymbol{\alpha}_\mathrm{c}\vert \boldsymbol{z})$ for a decreasing sequence of $\{\epsilon_t\}_{1\leq t\leq N_T}$ where $T$ is the maximum number of iterations \cite{silk2012optimizing}. The algorithm first generates an initial pool of $N$ particles that satisfy $\rho(\boldsymbol{z},\boldsymbol{z}^*)\leq\epsilon_1$ by randomly sampling from the prior $\pi(\boldsymbol{\alpha}_{\mathrm{c}})$. 
In the following iterations, successive distributions are randomly constructed by sampling from the previous population with probabilities $\{w^{(i,t-1)}\}_{\{1 \leq  i \leq N\}}$ where $w^{(i,t-1)}$ is the weight for the $i$th particle in iteration $t-1$. To filter and perturb the particles, we need a transition kernel. A transition kernel $\kappa^t$ is used to perturb the particles and find  $\boldsymbol{\alpha}_{\mathrm{c}}^{(i,t)}$'s. The new particle $\boldsymbol{\alpha}_{\mathrm{c}}^{(i,t)}$ is used to simulate $\boldsymbol{z}$ and if $\rho(\boldsymbol{z},\boldsymbol{z}^*)\leq\epsilon_t$ is satisfied, the particle is accepted. 
The process is repeated until $N$ particles are accepted. At iteration $t$, the ABC-SMC algorithm proposes parameters from the following distribution \cite{smart}
\begin{align}
q^\mathrm{t}&= \begin{cases}
\pi(\boldsymbol{\alpha}_\mathrm{c}), &\text{$t = 1$} \\
\sum_{j=1}^{N}w^{(j,t-1)}\kappa^t\Big(\boldsymbol{\alpha}_{\mathrm{c}}^{(i,t)} \big\vert \boldsymbol{\alpha}_{\mathrm{c}}^{(j,t-1)}\Big), & \text{ $ t > 1$},
\end{cases}
\end{align}
At each iteration, new weights are assigned to the particles, and in the next iteration the particles with larger weights become better represented in the population. The importance weights 
associated with an accepted population 
$\{\boldsymbol{\alpha}^{(i,t)}_{\mathrm{c}}\}_{\{1 \leq  i \leq N\}}$ are calculated as \cite{beaumont2009adaptive}:
\begin{align}\label{wbeamont}
w^{(i,t)}&= \begin{cases}
\frac{1}{N}, &\text{$t = 1$} \\
\frac{\pi\big(\boldsymbol{\alpha}_{\mathrm{c}}^{(i,t)}\big)}{\sum_{j=1}^{N}w^{(j,t-1)}\kappa^t\big(\boldsymbol{\alpha}_{\mathrm{c}}^{(i,t)} \big\vert \boldsymbol{\alpha}_{\mathrm{c}}^{(j,t-1)}\big)}. & \text{ $ t > 1$}
\end{cases}
\end{align}\\
The efficiency of ABC-SMC heavily relies on a proper choice of the perturbation kernel function $\kappa^{t}(\cdot \vert \cdot)$, the distance function, $\rho(\boldsymbol{z},\boldsymbol{z}^*)$, having a good prior distributions for the parameters, the threshold sequence  $\{\epsilon_t\}_{1\le t \le T}$, and  the weights of the particles in each iterations\cite{filippi2013optimality}. 
In this paper, we carefully assigned a distance function and focused on the adaptive weight and how to find the good prior distributions for the parameters which we do not know their initial values. These factors will be discussed below.
\subsection{Distance Function}
Choosing a summary statistic and distance metric which are sensitive to the parameters
of interest is a crucial step in parameter inference with ABC-SMC  \cite{smart}. 
 In this paper, we choose the following $L_2$ distance function based on numerical experiments:
\begin{align} \label{distance_function}
&\rho(\boldsymbol{z},\boldsymbol{z}^*) = \frac{1}{2K}\norm{\boldsymbol{z}-\boldsymbol{z}^*}_2,
\end{align}
where $\norm{\cdot}_2$ is the 2-norm of a vector.

\subsection{Probability Weight}

The probability weights of the particles allow the algorithm to search in the regions with high-probability
and to reject particles from low-probability regions of the parameter space \cite{ beaumont2009adaptive}. In \cite{beaumont2009adaptive, smart, filippi2013optimality}, the weights for the all particles at iteration $t=1$ are equal to $1/N$. In \cite{sisson2007sequential}, the weight considered based on the prior distribution for the parameters. However, since the particles are sampled randomly in the first iteration, the distance of the particles is different, and assigning an equal weight causes the algorithm to search around the particles which may be far from the optimal value. In this paper, we consider the weights based on the discrepancy of the particles. 

Let's consider the discrepancy vector at the first iteration as $\boldsymbol{\rho}^1 = [\rho_1^1, \rho_2^1,\cdots, \rho_N^1]$. Based on the discrepancies, for the particle which has a smaller discrepancy there is more probability of being close to the optimal, then the particle with smaller discrepancy should have a greater weight. Therefore, in this paper, we assigned the weight for the particle $i$ at the first iteration as follows:
\begin{align} \label{w1}
& w^{(i,1)} =  \frac{1}{\rho_i^1}.
\end{align}


For iteration $ 2 \leq  t  \leq T$, \cite{sisson2007sequential} used the prior distribution and a forward and a backward kernels to assign the weights for the particles, \cite{beaumont2009adaptive} improved the weights and used (\ref{wbeamont}) to calculate the weights of the particles.  But, in the cases where the prior distributions are not well known for the parameters, using the weight based on  (\ref{wbeamont}) is not a good choice. In this paper, for particle $i$ at iteration $t$ we use the prior information of the parameter, $\pi\big(\boldsymbol{\alpha}_{\mathrm{c}}^{(i,t)}\big)$, and the information of how much the particles are close to the optimal value, $\rho_i^t$. We defined a constant $\beta$ to make a trade-off between the prior information of the particles and their distances. This constant can help the algorithm not to  be stuck in search around the local optima. Then, the weight for $ 2 \leq  t  \leq T$ for particle $i$ is calculated as follow:
 \begin{align} \label{wt}
& w^{(i,t)} =  \beta \, \pi\big(\boldsymbol{\alpha}_{\mathrm{c}}^{(i,t)}\big) +(1-\beta) \,\frac{1}{\rho_i^t}.
\end{align}

\subsection{Adaptive Threshold Sequence}\label{tol}

To balance the computational efficiency and the accuracy of the posterior distribution, we define a threshold sequence: 
\begin{align}
    \mathcal{E} =\{\epsilon_1, \epsilon_2, \cdots, \epsilon_{T}\},
\end{align}
where $\epsilon_1>\epsilon_2> \cdots > \epsilon_{T}$. If the threshold is too large, too many proposed particles are accepted; if it is too small, the ABC  algorithm is not efficient since many proposed particles will be rejected \cite{smart}. 
Selecting it adaptively based on some quantile of the threshold in the previous iteration has better performance \cite{drovandi2011approximate}.
In this paper, we use the following threshold sequence scheme:

\begin{itemize}
\item We choose $\epsilon_1$ as the acceptance rate in the first iteration is equal to 0.5.  We run the simulation for $K_\mathrm{ini} = 2\, N$, and chose the median of the all discrepancy of $K_\mathrm{ini}$ simulations.   
\item For $\epsilon_{2:T-1}$, $\epsilon_{t+1}$ is calculated based on the $q\mathrm{th}$-percentile of the distribution of particle distances in iteration $t$.
\end{itemize}

\subsection{Prior Distribution Correction}\label{priorc}

An interesting and inexpensive feature of the proposed approach is based on the first step, in which we can estimate the parameters of the system, even we do not know the prior distribution of the parameters. For these cases, we add the prior correction at the first step of the algorithm.
For the parameters that we do not know good prior distributions, we consider a uniform distribution with very small lower and very large upper bounds. This step makes the proposed method robust to such deviation and makes it suitable for  cases in which one does not have appropriate prior knowledge about the true parameters. For instance, if the prior $\pi\big(\boldsymbol{\alpha}_{\mathrm{c}}) $ is  misspecified, it means that the true parameter is not contained in the support of $\pi\big(\boldsymbol{\alpha}_{\mathrm{c}})$. In the case of the power electronics application, for example, we do not know the impedance of the power supply.  
In this approach, we model the distribution    $p(\rho(\cdot) \vert \boldsymbol{\alpha}_\mathrm{c} )$ based on the parameter, i. e  for any value of input we calculate the discrepancy for $N_0$ simulation. 
Let consider the discrepancy for parameter $i$ as $\boldsymbol{\rho_0} = \{\rho_0^1, \rho_0^2, \cdots, \rho_0^{N_0}\}$. We consider the $N_\mathrm{p}$ smallest distance of the $N_0$ distances, and based on the $N_\mathrm{p}$ distances, we consider a Gaussian distribution for the prior distribution with the following mean and variance.
\begin{align}\label{priort}
    \mu = \min (\boldsymbol{\rho_0}), 
    \sigma^2=\frac{\sum_{i=1}^{N_\mathrm{p}}(\rho_0^i-\mu)^2}{N_\mathrm{p}}.
\end{align}
\section{Proposed ABC-SMC Algorithm}\label{sec:proposed_abc}
The proposed ABC-SMC algorithm is presented in Algorithm \ref{oursmc}. The ABC-SMC  algorithm will stop when the lowest threshold in the threshold sequence is less than the predefined smallest threshold or when a maximum number of $T$ iterations has been performed \cite{sisson2007sequential,beaumont2009adaptive}

\begin{algorithm}[!htb]
\captionsetup{font={small,sf,bf}, labelsep=newline}
  \caption{\small Adaptive ABC-SMC algorithm for estimating the posterior distribution of parameters $\boldsymbol{\alpha}_\mathrm{c}$ using $N$ particles, the prior distribution $\pi(\boldsymbol{\alpha}_\mathrm{c})$, given data $\boldsymbol{z}^*$. $\boldsymbol{\alpha}_{\mathrm{c}}^{(i,t)}$ is the parameter set for particle $i$ at iteration $t$.} 
  \small
\begin{algorithmic}[1]
\STATE \textbf{Find the prior distributions for the parameter based on (\ref{priort})}
\STATE \textbf{Set maximum number of iterations $T$ and set $\epsilon_1$ by section (\ref{tol})}
\STATE At iteration $t = 1$
\FOR{$1\leq i \leq N$}
\WHILE{$\rho(\boldsymbol{z},\boldsymbol{z}^*)>\epsilon_1$}
\STATE Sample $\boldsymbol{\alpha}_\mathrm{c}^*$ from the prior: $\boldsymbol{\alpha}_\mathrm{c}^*\sim \pi(\boldsymbol{\alpha}_\mathrm{c})$
\STATE Generate data $\boldsymbol{z}$ from $\boldsymbol{\alpha}_\mathrm{c}^*$: $\boldsymbol{z}\sim \mathrm{Model}(\boldsymbol{\alpha}_\mathrm{c}^*)$
\STATE \textbf{Calculate discrepancy $\rho(\boldsymbol{z},\boldsymbol{z}^*)$ based on (\ref{distance_function})}
\ENDWHILE
\STATE Set $\boldsymbol{\alpha}_{\mathrm{c}}^{(i,1)} \gets \boldsymbol{\alpha}^*_\mathrm{c}$ 
\STATE \textbf{Set $w^{(i,1)}$ based on (\ref{w1})}  
\ENDFOR 
\STATE \textbf{Generate Gaussian perturbation kernel $\kappa^2 = \mathcal{N}( \tilde{\boldsymbol{\alpha}}_{\mathrm{c}}^1, \Gamma^1)$} 
\STATE \textbf{Determine $\epsilon_2$ based on section (\ref{tol})}
\STATE At iteration $t>1$
\FOR{$2\leq t \leq T$}
\FOR{$1\leq i \leq N$}
\WHILE{$\rho(\boldsymbol{z},\boldsymbol{z}^*) > \epsilon_{t}$}
\STATE Sample $\boldsymbol{\alpha}_\mathrm{c}^*$  from the previous population                          $\{{\boldsymbol{\alpha}}_{\mathrm{c}}^{(i,t-1)}\}_{\{1 \leq  i \leq N\}}$ with probabilities $\{w^{(i,t-1)}\}_{\{1 \leq  i \leq N\}}$ and perturb them to obtain $\boldsymbol{\alpha}_\mathrm{c}^{**}\sim \kappa^t( \boldsymbol{\alpha}_{\mathrm{c}}^t, 2\,\Gamma^{t-1})$
\STATE Generate data $\boldsymbol{z}$ from $\boldsymbol{\alpha}_\mathrm{c}^{**}:\boldsymbol{z} \sim \mathrm{Model}(\boldsymbol{\alpha}_\mathrm{c}^{**})$
\STATE \textbf{Calculate discrepancy $\rho(\boldsymbol{z},\boldsymbol{z}^*)$ based on (\ref{distance_function})}
\ENDWHILE
\STATE Set $\boldsymbol{\alpha}_\mathrm{c}^{t}\gets\boldsymbol{\alpha}_\mathrm{c}^{**}$
\STATE \textbf{Calculate $w^{(i,t)}$ based on (\ref{wt})}
\ENDFOR
\STATE \textbf{Generate Gaussian perturbation kernel $\kappa^{t+1} =  \mathcal{N}( \boldsymbol{\alpha}_{\mathrm{c}}^t, 2\,\Gamma^{t-1})$}
\STATE \textbf{Determine $\epsilon_{t+1}$ based on section (\ref{tol})}
\ENDFOR
\end{algorithmic}\label{oursmc}
\end{algorithm}

\section{Simulation Results} \label{result}

The model of the DC-DC buck converter is based on TPS40200EVM‐002 model built in Matlab/Simulink and the proposed algorithm is implemented in Python. 
All tests are performed on a desktop PC with Intel(R) Core(TM) i7-8700 and 8-GB RAM. 

\subsection{Parameter setting}
In this paper, in all simulations, we consider $q=0.75$ for choosing the thresholds. Based on the simulations results, we chose the $\beta = 0.4$. We consider a  Gaussian distribution kernel same as \cite{beaumont2009adaptive}. We set the maximum number of iteration as $T=10$. 

\subsection{Adaptive Weight}
\begin{figure}[!htb]
	\centering
	\subfloat[]{\includegraphics[height=1.8in, width=2.8in]{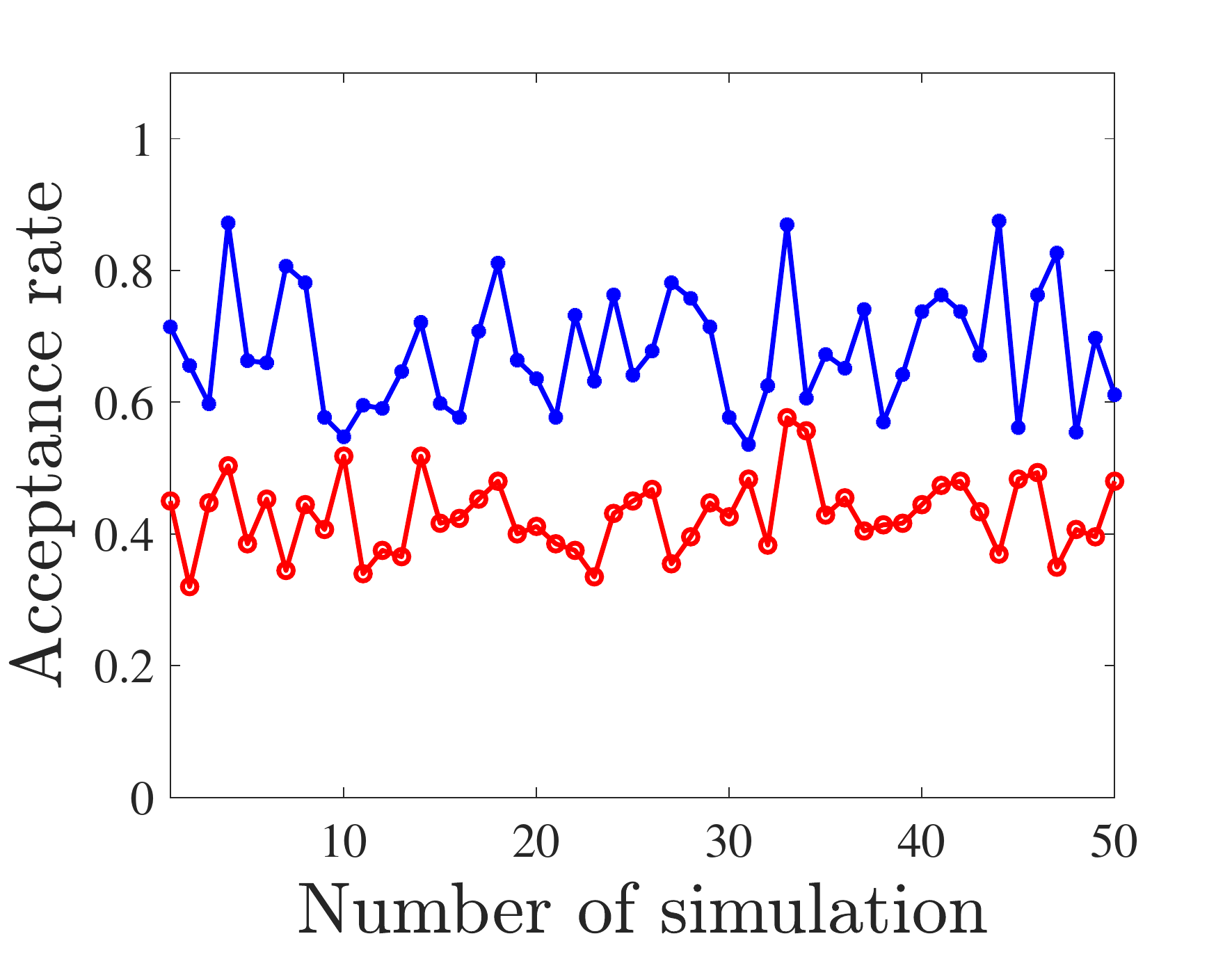}\label{secondit}}
	\\
	\subfloat[]{\includegraphics[height=1.8in, width=2.8in]{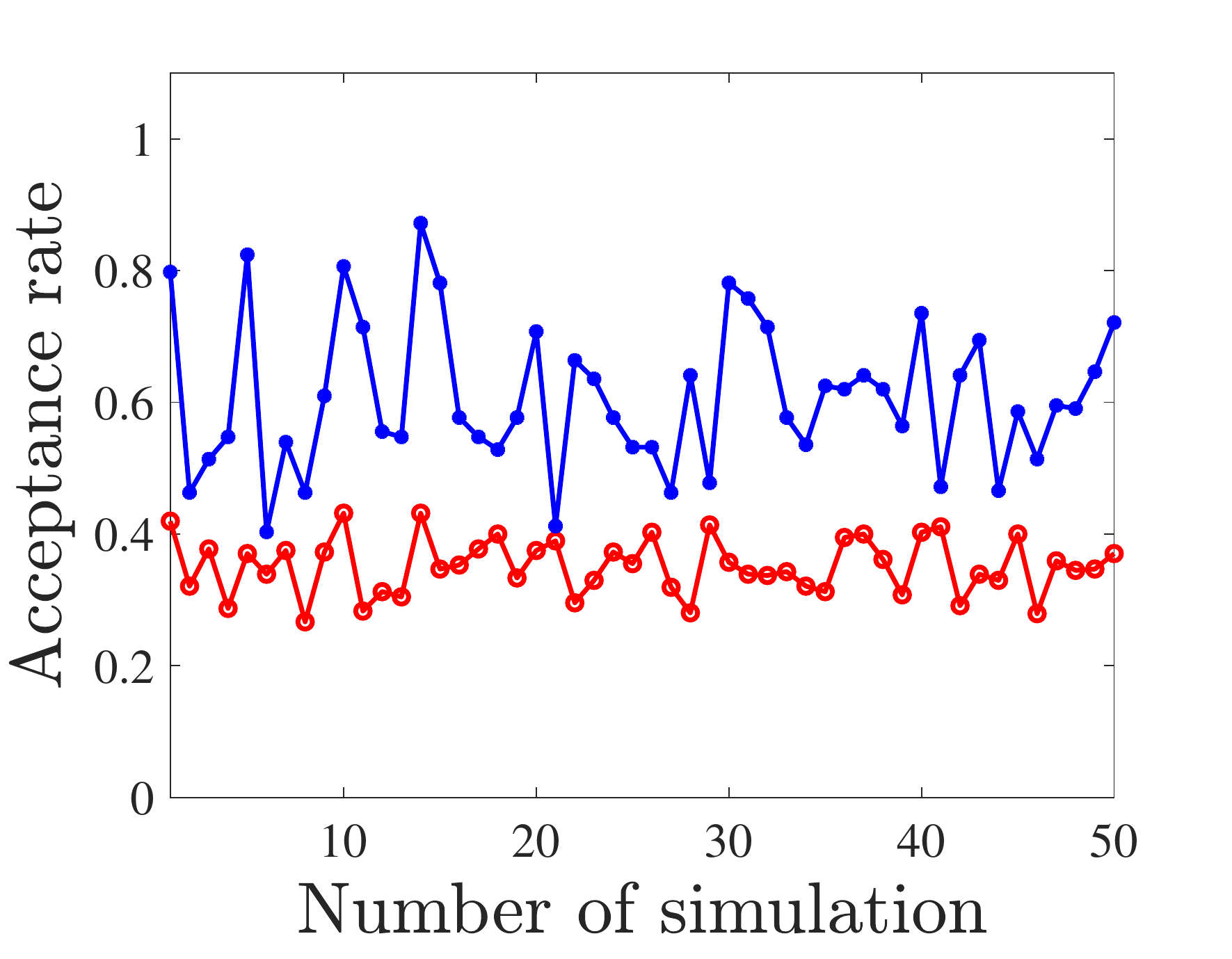}\label{whole}}
	\captionsetup{justification=raggedright,singlelinecheck=false}
	\caption{Acceptance rates for fifty different simulations (a) $t=2$ and (b) whole iterations.   \textcolor{red}{$\circ$} Weight in \cite{beaumont2009adaptive, toni2009approximate} \textcolor{blue}{$\ast$} Proposed weight.}
\end{figure}

For comparing different methods, we use the acceptance rate defined as follow:
\begin{align}
    \text{acc} = \frac{N}{N_\mathrm{s}},
\end{align}
where $N$ is the number of the particles  used in the algorithm, and $N_\mathrm{s}$ is the total number of simulation during each iteration. At first, we compare the $\text{acc}$ at iteration 2 with the $w_\mathrm{e}$ which is used in \cite{beaumont2002approximate, beaumont2009adaptive, toni2009approximate}. Fig. \ref{secondit} shows the acceptance rate for fifty independent simulations with the proposed weight and the $w_\mathrm{e}$. As it can be seen, the proposed weight has greater acceptance rate for all the simulations.  

Fig. \ref{whole} shows the acceptance rate for whole iterations of fifty different simulations. It can be seen that the algorithm can find the posterior distributions with less number of simulations with compare to the other methods.

\subsection{Critical Parameter Identification}
\vspace{0.2cm}
From the trajectory sensitivity in Section \ref{sen}, we consider eight critical parameters which have the main impacts on the outputs.  The top eight
parameters and their normalized sensitivities are listed in Table \ref{sensitivity}.

\begin{table}[htb!]
\centering
\renewcommand{\arraystretch}{0.9}
\captionsetup{labelsep=space,font={footnotesize,sc}}
\caption{Sensitivity Analysis of the Parameters}
\begin{tabular}{c|c}
\hline\hline \xrowht{16pt}
\textbf{Parameter} & \textbf{Sensitivity} \\ \hline
$L$   &   1         \\ \hline
$C_2$   &   0.81          \\ \hline
      $R_\mathrm{s}$   &   0.75         \\ \hline
        $L_\mathrm{s}$  &    0.68         \\ \hline
         $C_\mathrm{in}$ &     0.65        \\ \hline
        $R_\mathrm{c1}$  &      0.61       \\ \hline
         $R_\mathrm{c2}$ &      0.45       \\ \hline
         $C_1$ &     0.41        \\ \hline\hline
\end{tabular}\label{sensitivity}
\end{table}
\subsection{Calibration of Buck Converter}\label{model}

In this paper, we consider the non-isolated buck converter. Its topology is shown in Fig. \ref{fig:buck}.
At first, we get the outputs of the Buck converter with a set of parameters, $\boldsymbol{\alpha_{\mathrm{c}}^\mathrm{True}}$, which we know their values. We consider the outputs of the model regarding the $\boldsymbol{\alpha_{\mathrm{c}}^\mathrm{True}}$ as $\boldsymbol{z}^*$. 
For the resistance of the power supply, we assume that we do not the prior distribution for it. Then, we consider a uniform distribution as $\mathcal{U}(0, 10000)$ to consider all uncertainties. Then, by the simulation based on section \ref{priorc} we found that the prior distribution can be considered as a Gaussian distribution as $\mathcal{N}(0.5, 10)$. 
For the other parameters, we consider the uniform distribution as the prior distributions for the parameters and estimate their values. We consider the mean values of the parameter as 20\% percent greater than the true value to consider the uncertainties. 
We choose the lower/upper bounds of the uniform prior distributions for the parameters as a very small number and very large number. Table \ref{table:es} shows the prior distributions, the estimated values, and the estimation errors. 
It is seen that the proposed method can accurately estimate the parameters under a uniform prior distribution with a small/large for lower/upper bounds. 

\begin{figure}[!htb]
	\centering
	\subfloat[]{\includegraphics[height=1.8in, width=2.8in]{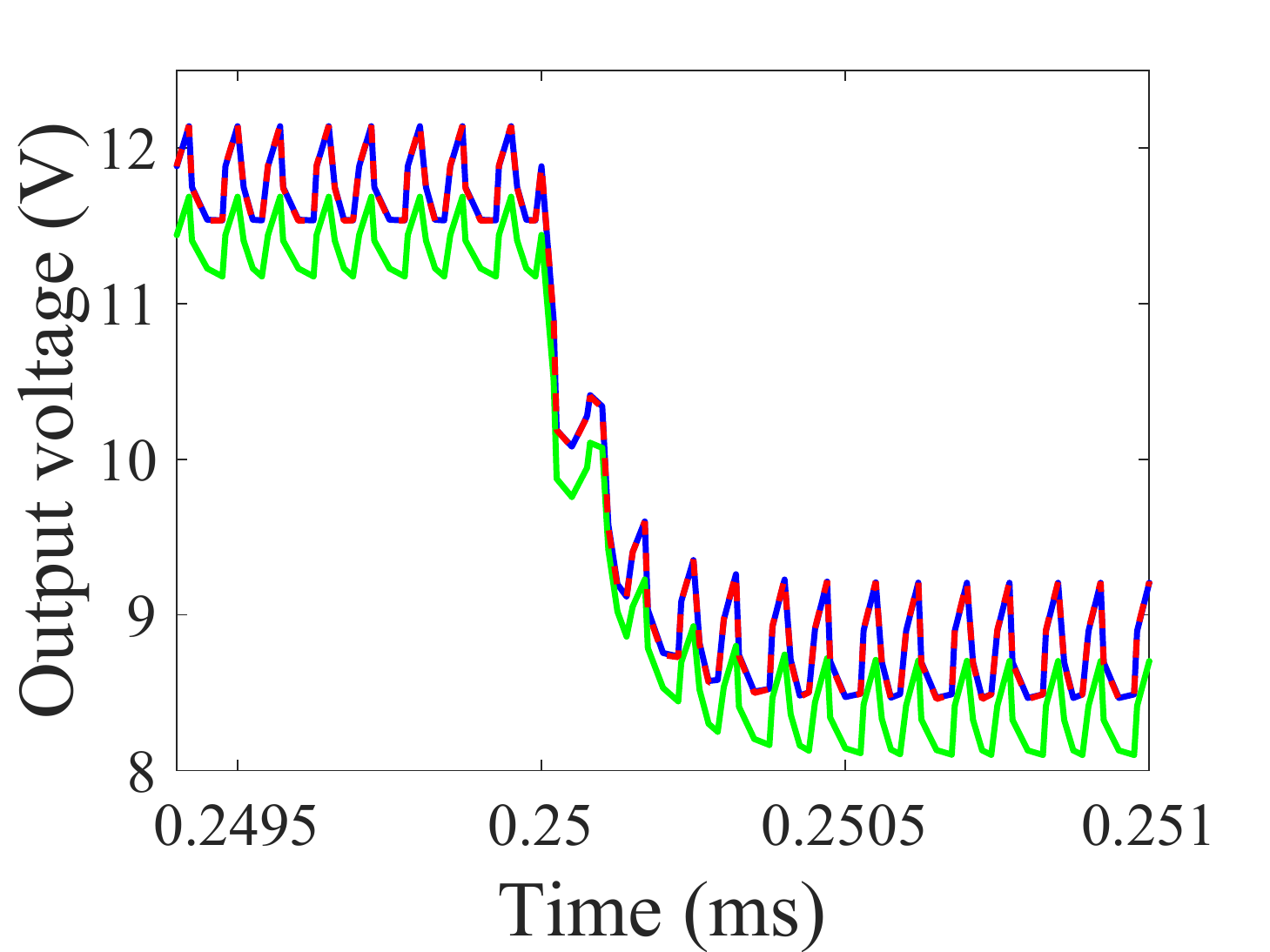}\label{vtr}}
	\\
	\subfloat[]{\includegraphics[height=1.8in, width=2.8in]{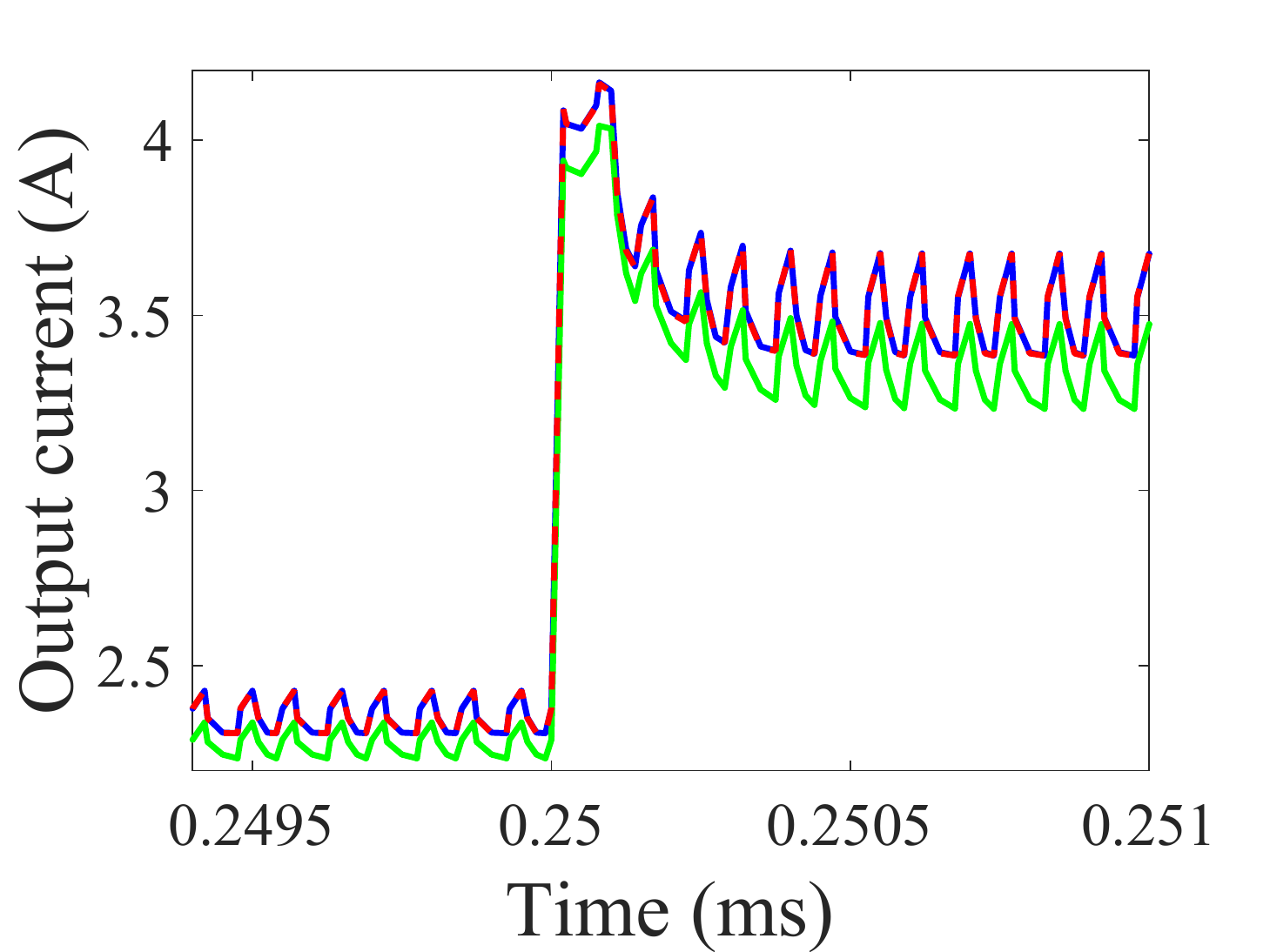}\label{itr}}
	\captionsetup{justification=raggedright,singlelinecheck=false}
	\caption{Buck converter operating under transient
condition. (a) Output voltage; (b) Output current.  \textcolor{blue}{\textbf{\textemdash}} Measurements; \textcolor{green}{\textbf{.-}}  converter before calibration; \textcolor{red}{\textbf{- -}}  converter after calibration.}
	\label{fig:trn}
\end{figure}

\begin{figure}[!htb]
	\centering
	\subfloat[]{\includegraphics[height=1.7in, width=2.8in]{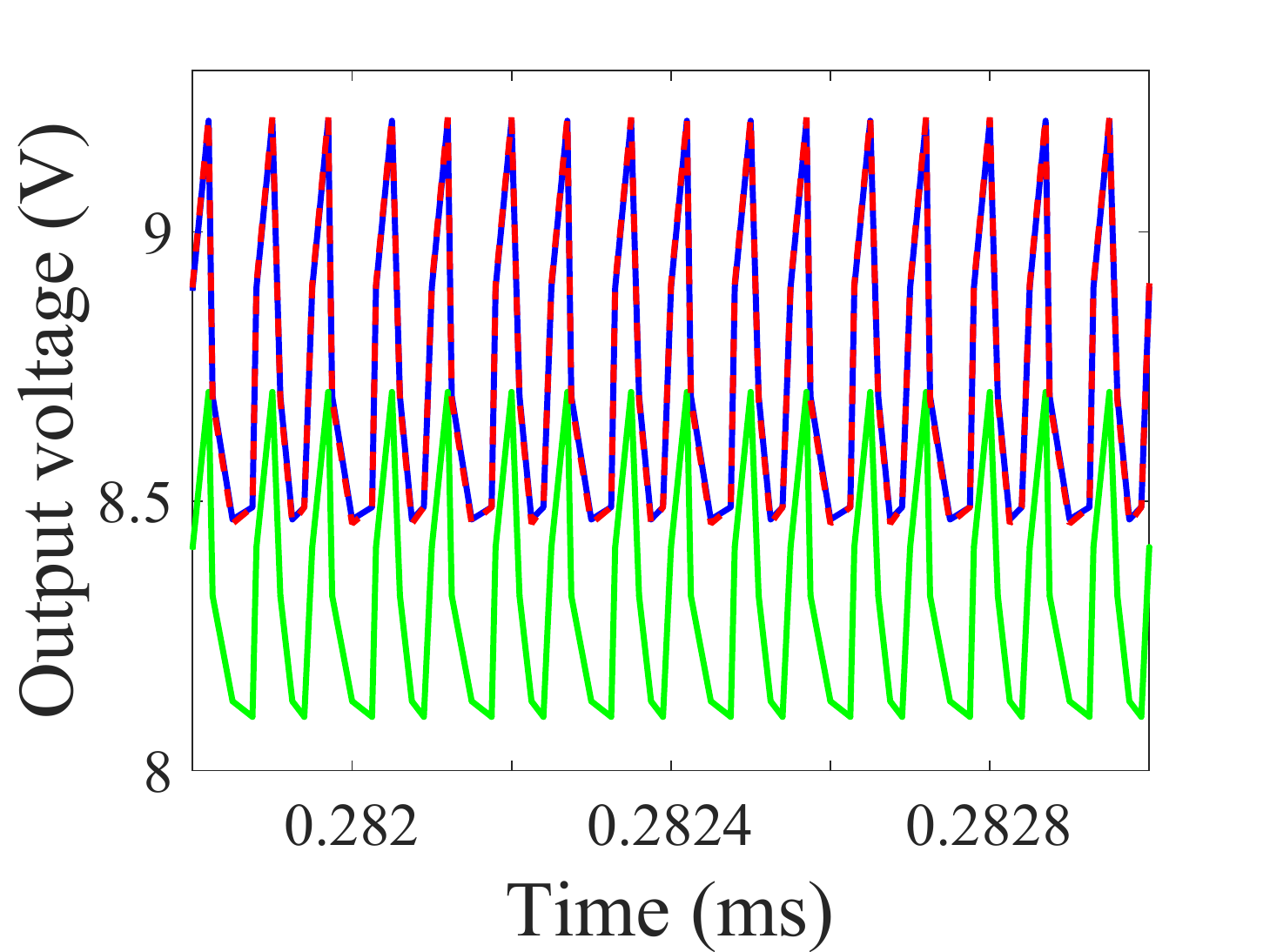}\label{vs}}
	\\
	\subfloat[]{\includegraphics[height=1.7in, width=2.8in]{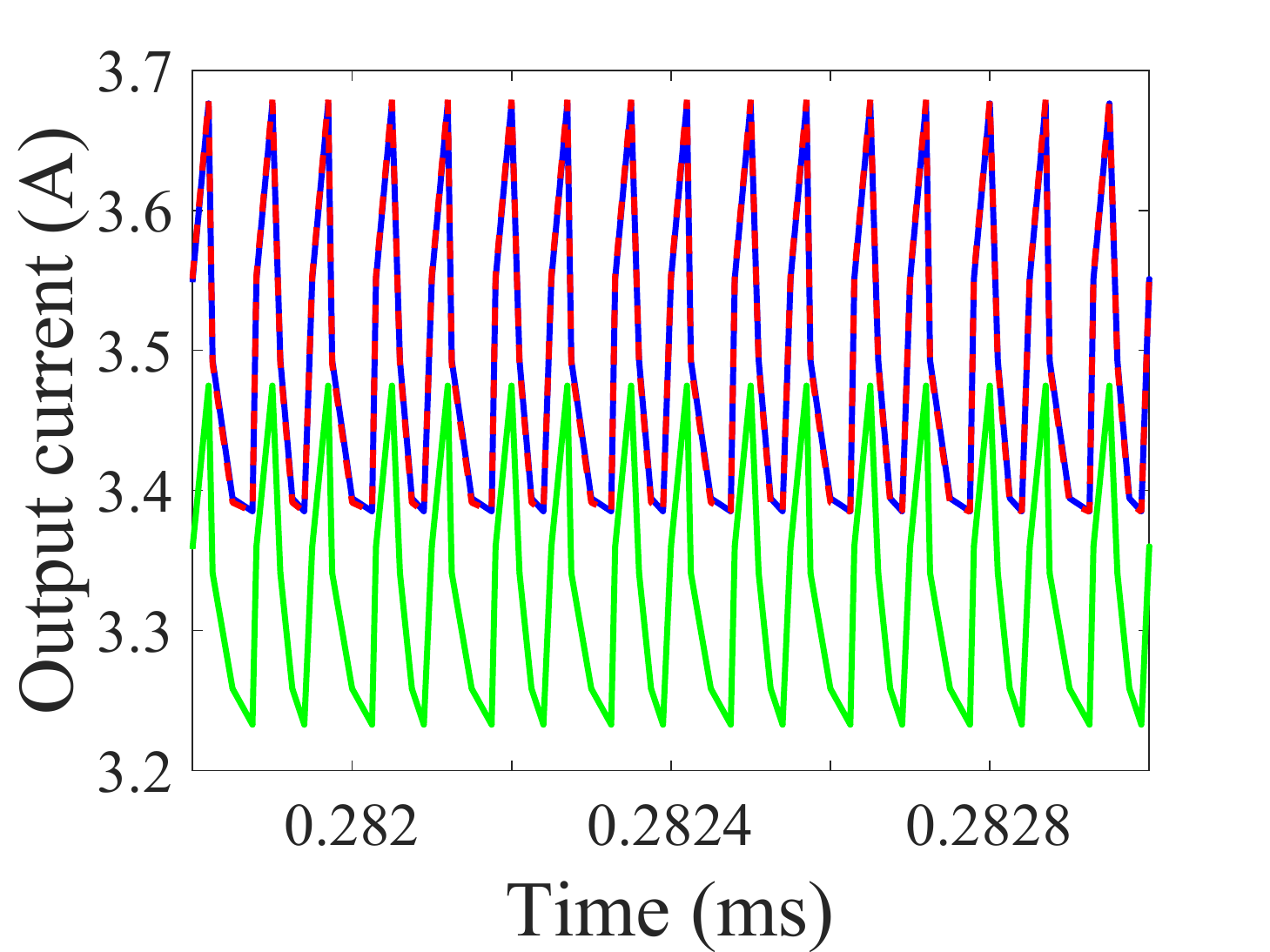}\label{Is}}
	\captionsetup{justification=raggedright,singlelinecheck=false}
	\caption{Buck converter operating under steady state 
condition. (a) Output voltage; (b) Output current.  \textcolor{blue}{\textbf{\textemdash}} Measurements; \textcolor{green}{\textbf{.-}}  converter before calibration; \textcolor{red}{\textbf{- -}}  converter after calibration..}
	\label{st}
\end{figure}

To analyze the performance of the converter with the estimated parameters under the transient and steady-state conditions, we change the load at the output. Fig. \ref{fig:trn} shows the performance of the converter under the transient condition, and Fig. \ref{st} shows the performance of the converter in the steady-state condition. As can be seen, the output of the converter with the estimated parameters is very close to the measurements.

\begin{table}[htb]
\renewcommand{\arraystretch}{1.1}
\captionsetup{labelsep=space,font={footnotesize,sc}}
\caption{Actual and Identified Values of the Parameters of the Buck Converter.}

\footnotesize
\centering
\renewcommand{\arraystretch}{1.7}
\begin{tabular}{|c|c|c|c|c|}
\hline\hline
 \textbf{Parameter} & \tabincell{c}{\textbf{True} \\\textbf{value}} & \tabincell{c}{\textbf{Prior} \\ \textbf{distribution}}   & \tabincell{c}{\textbf{Estimated} \\ \textbf{value}}    & \textbf{ \% Error}   \\ \hline\hline
  $L$&  \SI{33}{\mu\henry} &$\mathcal{U}$(0,   1.3\,m)  &\SI{32.56}{\mu\henry}  & 1 \\ \hline
  $C_2$ & \SI{100}\mu\si{{\farad}} & $\mathcal{U}$(0,   0.01) & \SI{99.3}\mu\si{{\farad}} & 0.7 \\ \hline
  $R_\mathrm{s}$ &  \SI{0.16}{\ohm}  & $\mathcal{N}$(0.5,   8) & 0.16 & 0 \\ \hline
 $L_\mathrm{s}$ &  \SI{0.40}{\micro\henry}  & $\mathcal{U}$(0,   5) & \SI{0.40}{\micro\henry} & 0 \\ \hline
 $C_\mathrm{in}$ &  \SI{100}\mu\si{{\farad}} &$\mathcal{U}$(0,   0.01)  &  \SI{99.2}\mu\si{{\farad}} & 0.8 \\ \hline
 $R_\mathrm{c1}$ &\SI{65}{\mohm}  &$\mathcal{U}$(0,   0.5)  & \SI{64.8}{\mohm} &  0.03\\ \hline
 $R_\mathrm{c2}$ & \SI{300}{\mohm} & $\mathcal{U}$(0,   1) & \SI{300.8}{\mohm} & 0.20 \\ \hline
$C_1$ & \SI{100}\mu\si{{\farad}} & $\mathcal{U}$(0,   0.01) & \SI{100}\mu\si{{\farad}} & 0 \\ \hline\hline


\end{tabular}\label{table:es}
\end{table}

\section{Conclusion}\label{conclusion}

In this paper, we proposed a parameter calibration method
for DC-DC buck power converter based on an adaptive Approximate Bayesian Computation with sequential Monte Carlo  sampler (ABC-SMC) approach. We developing the ABC-SMC algorithm by proposing a novel and straightforward weight scheme. The proposed algorithm tested on a DC-DC converter with its parasite and passive elements of the converter. Test results show that the proposed approach can find the exact values of the parameters for a converter by considering the passive and parasite components. We also analyze the steady-state and transient performance of the converter with the estimated parameters. The results show the great performance of the algorithm. However, in this work we only implement the method on the simulation based converter, and we will further test the proposed algorithm on a real case with its controllers.

\bibliographystyle{IEEEtran}	
\bibliography{Refrences}
\end{document}